\def\fun#1#2{\lower3.6pt\vbox{\baselineskip0pt\lineskip.9pt
  \ialign{$\mathsurround=0pt#1\hfil##\hfil$\crcr#2\crcr\sim\crcr}}}
\relax
\documentstyle[12pt]{article}
\advance\textheight by \topskip
\begin{document}
\vspace{-0.2cm}
\begin{flushright}
UG-3/94\\
SU-ITP-94-11\\
QMW-PH-94-15\\
hep-th/9406009
\end{flushright}
\vspace{-0.5cm}

\begin{center}
{\large\bf BLACK-HOLE-WAVE DUALITY\\
\vskip .6 cm
IN STRING THEORY }\\
\vskip 2 cm

{\bf Eric Bergshoeff ${}^a$\footnote{ E-mail address:
{\tt bergshoe@th.rug.nl}},
Renata Kallosh ${}^b$\footnote {  E-mail address:
{\tt kallosh@physics.stanford.edu}}
and Tom\'as Ort\'{\i}n${}^{c}$\footnote{
E-mail address: {\tt ortin@qmchep.cern.ch}}}
\vskip 0.05cm
${}^a$ Institute for Theoretical Physics, University of Groningen\\
Nijenborgh 4, 9747 AG Groningen, The Netherlands\\
\vskip 0.4truecm
${}^b$ Department of
Physics,
 Stanford University\\
Stanford   CA 94305, USA\\
\vskip .4truecm
 ${}^{c}$ Department of
Physics, Queen Mary and Westfield College, \\
Mile End Road, London E1
4NS, U.K.
\end{center}

\vskip 1.0 cm
\centerline{\bf ABSTRACT}
\begin{quotation}

\noindent Extreme 4-dimensional dilaton black holes embedded into
10-dimensional geometry are shown to be dual to the gravitational
waves in string theory. The
corresponding gravitational waves are the generalization of pp-fronted
waves, called supersymmetric string waves. They are given by  Brinkmann
metric  and the two-form field, without a dilaton. The non-diagonal part of the
metric of the
dual partner of the wave together
with the two-form field correspond to the vector field in 4-dimensional
geometry of the charged extreme black holes.

\end{quotation}

\newpage

1. In this paper we continue our investigation of  black holes and
gravitational waves. From the point of view of General Relativity those are
very different geometries. However the string theory brings in  a completely
different
concept of  ''equivalent''  background geometries. It was understood  some
time ago  \cite{Ho4} that the pp-waves are dual to the fundamental strings
\cite{Da1}.
The corresponding duality transformation, which is known as sigma-model
duality transformation  \cite{BUS},  has a very particular property: it changes
the value of the dilaton field $e^{2\phi}$ by the $g_{xx}$-component of the
metric, where $x$ is some direction on which all fields are independent. In our
recent paper \cite{BEK} we have established an analogous  dual relation between
  more general solutions of the effective equations of the critical ($d=10$)
superstring theories.
The purpose of this letter is to show that some solutions, which can be
obtained
 by dual rotation from  a particular case of supersymmetric string waves (SSW)
\cite{BKO} are identified as supersymmetric extreme charged dilaton
black holes upon Kaluza-Klein dimensional reduction to $d=4$.

The first known to us example of a deep relation between gravitational waves
and black-hole type solutions was given by Gibbons \cite{GW}. He  has
observed that the 5-dimensional pp-waves upon Kaluza-Klein
dimensional reduction to $d=4$ are equivalent to a singular limit of
electrically charged black hole. Those black holes have the scalar field
 $g_{55}=\sigma$ coupling to the
 vector field of the form $e^{2\sqrt 3\sigma}F^2$. The electric solutions are
 related via  electric-magnetic duality to monopoles. Thus this
example has shown the dual relation between gravitational waves and
monopoles.

We will use the sigma-model duality of the string theory and relate solutions
of 4-dimensional and 10-dimensional effective actions of string theory.
We will limit ourselves by keeping only one scalar field, the fundamental
dilaton. The method which we develop here may give many other
interesting relations for the class of solutions which will include more
fields of the string theory.

2. We  consider the zero slope limit of the effective string action. This limit
corresponds to 10-dimensional $N=1$ supergravity. The Yang-Mills multiplet
will appear in the first order of $\alpha'$ string corrections.
The SSW \cite{BKO} in $d=10$
 are given by the Brinkmann metric \cite{Br1}
and the following 2-form
\begin{eqnarray}
\label{SSW}
ds^2 &=& 2 d \tilde u d \tilde v + 2A_M d\tilde x^M d\tilde u -
 \sum_{i=1}^{i=8} d\tilde x^id \tilde x^i\ ,\nonumber \\
\label{eq:SSW2}
B &=& 2 A_M d\tilde x^M \wedge d\tilde u\ , \qquad
A_{v} =0 \ ,
\end{eqnarray}
where $i=1,\dots , 8, \; M = 0, 1, \dots , 8, 9$ and we are using the
 following notation for the
10-dimensional coordinates
$x^M = \{\tilde u, \tilde v, \tilde x^i\}$. We have put the tilde over
the 10-dimensional coordinates, since we will have to compare
the original 10-dimensional
configuration with the  4-dimensional one, embedded into the 10-dimensional
space. A rather non-trivial identification of coordinates, describing these
solutions
will be required later.

The equations that $A_u(\tilde x^i)$ and $A_i(\tilde x^j)$ have to satisfy are:
\begin{equation}
\label{eq:Lapl}
\triangle A_u = 0\ , \hskip 1.5truecm \triangle\partial^{[i}A^{j]} =
0\ ,
\end{equation}
where the Laplacian is taken over the transverse directions only.

 Sigma-model duality transformation \cite{BUS} defines the changes in the
metric, 2-form field $B_{\mu\nu}$ and in the dilaton field $e^{2\phi}$ :
\begin{eqnarray}
 g_{xx} ' & =& 1/g_{xx}\ , \qquad  g_{x\alpha} ' =
B_{x\alpha}/
g_{xx}\ , \nonumber\\
  g_{\alpha\beta}' & =& g_{\alpha\beta} -
(g_{x\alpha}g_{x\beta} -
B_{x\alpha}B_{x\beta})/g_{xx}\ , \nonumber\\
  B_{x\alpha} ' & =& g_{x\alpha}/g_{xx}\ , \qquad
	  B_{\alpha\beta} ' = B_{\alpha\beta} +2
g_{x[\alpha}
B_{\beta]x}/g_{xx}\ , \nonumber\\
 \phi  '& =& \phi - {1\over 2} \log |g_{xx}| \ .
\label{bus}\end{eqnarray}
This transformation is defined for configurations with a non-null
Killing vector
in the $x$-direction.
The string theory considers such configurations as equivalent under
the condition
that the $x$-direction is compact.

A straightforward application of the sigma-model duality
transformations
given in (\ref{bus}) on the  SSW solution given in
eq.~(\ref{SSW})
leads to the following new supersymmetric solution of the
zero slope limit equations of motion:
\begin{eqnarray}
\label{dualwave}
ds^2 &=&  2e^{2\phi}\bigl \{ d\tilde ud\tilde v +  A_i d\tilde ud\tilde x^i
\bigr \} - \sum_{i=1}^{i=8} d\tilde x^id\tilde x^i\ ,\nonumber\\
B &=& -2 e^{2\phi} \bigl
\{ A_u d\tilde u \wedge d\tilde v +  A_id\tilde u \wedge d\tilde x^i \bigr \}\
,\\
e^{-2\phi}  &=& 1 - A_u \ ,\nonumber
\end{eqnarray}
where as before, the functions $A_{M} = \{A_u= A_u (\tilde x^j), A_v=0,
 A_i = A_i (\tilde x^j)\}$
satisfy equations (\ref{eq:Lapl}).
We called this solution generalized fundamental strings \cite{BEK},
since at that time we had in mind only  a subclass of these solutions
 which depend on the 8-dimensional transverse coordinates and has an
interpretation of a macroscopic string. However, if we  assume that the
 functions $A_{M} $ do not depend on all 8 transverse coordinates, but
only on  part of them, the name given to this class of solutions is not
appropriate anymore. To avoid possible confusion we would call the generic
solutions given in eq. (\ref{dualwave})  the dual partner of
the SSW\footnote{Indeed,  we are going to show that the  dual partner of the
SSW corresponds, in particular,  to the lifted black holes,  when
the functions $A_M$ depend only on the coordinates of the 3-dimensional
space.
Therefore we want to stress that   the dual partner of the SSW include  more
general configurations than strings.}.

We can make the following particular choice of the vector
function $A_M$. First of all  these functions will depend only on 3 of
the transverse coordinates, $\tilde x^1, \tilde x^2, \tilde x^3$, which
will eventually correspond to
our 3-dimensional
space. Secondly, we choose one of $A_i$ e. g. $A_4$ to be related to $A_u$.
\begin{equation}
A_u = - { \mu \over \rho}, \qquad  A_4 = \xi A_u \ , \qquad A_1=A_2=A_3=A_5 =
\dots = A_8=0
\end{equation}
where $\rho^2 = \sum_{i=1}^{i=3} \tilde x^i  \tilde  x^i \equiv \vec x ^2$
and $\mu$ is a constant.
We will specify the constant
$\xi$ later.
 Note that eqs. (\ref{eq:Lapl}) are solved outside $\rho=0$
\footnote{In order to solve the equations
(\ref{eq:Lapl}) everywhere,
it is understood that a source term at $\rho =0$, representing an unknown
object, perhaps a six-brane,
has to be added to these equations.  We hope that this point can be  worked
out in an
analogy with the combined action for the macroscopic fundamental string,
where the source term comes from the sigma-model action, see eqs. (3,1) -
(3,3) in \cite{Da1}.}. We get
\begin{eqnarray}
ds^{2} & = &
2e^{2\phi}\{d\tilde{u}d\tilde{v}+\xi (1-  e^{-2\phi})d\tilde{x}^{4}d\tilde{u}\}
- \sum_{i=1}^{i=8} d\tilde{x}^{i}d\tilde{x}^{i}\, ,
\nonumber \\
B & = & -2e^{2\phi} (1-e^{-2\phi})\{d\tilde{u}\wedge
d\tilde{v}+ \xi d\tilde{u}\wedge d\tilde{x}^{4}\}\, ,
\nonumber \\
e^{-2\phi}  &=& 1 + { \mu \over \rho}
 \ .
\label{eq:new}\end{eqnarray}

We  perform the coordinate change
\begin{equation}
\hat{x}=\tilde{x}^{4} + \xi \tilde{u}\,  , \qquad
\hat  v = \tilde v  + \xi  \tilde x^4 \ .\end{equation}
We also shift $B$ on a constant value, since equations of motion depend on
$H=dB$ only.

 The dual wave solution (\ref{eq:new}) takes the form
\begin{eqnarray}
ds^{2} & = &
2e^{2\phi}d\tilde{u}d\hat {v}+\xi^{2}d\tilde{u}^{2}-d\hat{x}^{2}
-\sum_{i=1}^{i=3} d\tilde{x}^{i}d\tilde{x}^{i} -
\sum_{i=5}^{i=8} d\tilde{x}^{i}d\tilde{x}^{i}\,
, \nonumber \\
B & = & -2e^{2\phi}d\tilde{u}\wedge d\hat {v}\, ,
\nonumber \\
e^{-2\phi}  &=& 1 + { \mu \over \rho} \ .
\end{eqnarray}
When $\xi^2 = -1$ we have
\begin{eqnarray}
ds^{2} & = &
2e^{2\phi} d\hat{v} d\tilde{u} - d\tilde{u}^{2}-d\hat{x}^{2}
-\sum_{i=1}^{i=3} d\tilde{x}^{i}d\tilde{x}^{i} -
\sum_{i=5}^{i=8} d\tilde{x}^{i}d\tilde{x}^{i}\,
, \nonumber \\
 B & = & -2e^{2\phi}d\tilde{u}\wedge d\hat {v}\, ,
\nonumber \\
e^{-2\phi}  &=& 1 + { \mu \over \rho} \ .
\label{partner}\end{eqnarray}
We can identify this particular dual partner of the SSW solution with
the uplifted dilaton black hole if we make the following identification
of coordinates
\begin{eqnarray}
t & = & \hat{v} =  \tilde v +\xi  \tilde x^4\, ,
\nonumber \\
x^{4} & = & \tilde{u}\, ,
\nonumber \\
x^{9} & = & \hat{x}= \tilde{x}^{4} + \xi \tilde{u}\, ,
\nonumber \\
x^{1,2,3,5,\dots,8} & =  & \tilde{x}^{1,2,3,5,\dots,8}\, .
\end{eqnarray}
Our dual wave becomes
\begin{eqnarray}
ds^{2} & = &
2e^{2\phi} dt d x^4 - \sum _4^9
d x^{i}d x^{i} -
d \vec x^2\,
, \nonumber \\
B & = & -2e^{2\phi}dx^4\wedge dt\, ,
\nonumber \\
e^{-2\phi}  &=& 1 + { \mu \over \rho} \ .
\label{10bh}\end{eqnarray}
This is an extreme electrically charged 4-dimensional black hole, which is
embedded into 10-dimensional
geometry in stringy frame, as we are going to explain in the next section.

3. The embedding of the 4-dimensional bosonic solutions of the effective
superstring action into 10-dimensional geometry is not unique, in general
\footnote{We are grateful to E. Witten for  attracting our attention to
this problem.}. There are different ways to identify the vector field of the
charged black hole in 4 with the non-diagonal component of the metric
in the extra dimensions as well as with the 2-form field. Also the
identification of the 4-dimensional dilaton with the fundamental
10-dimensional dilaton and/or with some components of the metric in
the extra dimension is possible.

However the identification of  the 4-dimensional solution with the
10-dimensional one becomes  unique under the conditions
that the supersymmetric embedding for both solutions is identified.
Dimensional reduction of $N=1$ supergravity down to $d=4$ has
been studied by Chamseddine \cite{Cham} in canonical geometry. We
are working in stringy metric and also in slightly different notation.
In a subsequent publication we will present a detailed derivation of the
compactification of the bosonic part of the effective action of the
10-dimensional string theory
which is consistent with supersymmetry \cite{BKO2}.
Here we are interested in the relation between the  extreme dilaton
charged black
holes,
which have unbroken supersymmetry \cite{US} when imbedded into $d=4,
N=4$ supergravity\footnote{We do not know at present, whether the
embedding of these black holes into other theories, including the Abelian
part of Yang-Mills multiplet, will also correspond to some unbroken
supersymmetries.} and the corresponding 10-dimensional supersymmetric
configuration.

We start with the zero slope limit of the effective 10-dimensional
superstring action. The bosonic part of the action is
\begin{equation}
S=\frac{1}{2}\int d^{10}x
e^{-2 \phi}\sqrt{- g} \, [ -R+
4 (\partial\phi )^{2}-\frac{3}{4}H^{2}]\, ,
\label{eq:actionD1}
\end{equation}
where the 10-dimensional fields are the metric, the axion and the dilaton.

We want to make connection with the bosonic part of $N=4$, $d=4$ action.  In
this
particular case we are interested in compactifying $6$ space-like
coordinates. All fields are assumed to be independent of
six compactified dimensions. According to Chamseddine \cite{Cham}
dimensional
reduction of $N= 1, \, d=10$ supergravity to $d=4$
gives $N=4$ supergravity coupled to  6 matter multiplets. We are interested
here
only in dimensional reduction to $N=4$ supergravity without matter multiplets.

Let us first reduce from $d=10$ to $d=5$ by trivial dimensional reduction,
when we do not keep the non-diagonal
components of the metric and 2-form field.
We denote the
$10$-dimensional fields by un upper index ${}^{(10)}$ and the $5$-dimensional
fields by a hat.  The $10$-dimensional indices are capital letters
$M,N=0,\ldots,9$, the $5$-dimensional indices will carry a hat
$\hat{\mu},\hat{\nu}=0,\ldots,4$, and the compactified dimensions will
be denoted by capital $I$'s and $J$'s, $I,J=5,\ldots,9$.  We take the
$d=10$ fields to be related to the $d=5$ ones by
\begin{eqnarray}
g^{(10)}_{\hat{\mu}\hat{\nu}} & = & \hat{g}_{\hat{\mu}\hat{\nu}}\, ,
\nonumber \\
g^{(10)}_{I\hat{\nu}} & = & 0\, ,
\nonumber \\
g^{(10)}_{IJ} & = & \eta_{IJ}=-\delta_{IJ}\, ,
\nonumber \\
B^{(10)}_{\hat{\mu}\hat{\nu}} & = & \hat{B}_{\hat{\mu}\hat{\nu}}\, ,
\nonumber \\
B^{(10)}_{I\hat{\nu}} & = & 0\, ,
\nonumber \\
B^{(10)}_{IJ} & = & 0\, ,
\nonumber \\
\phi^{(10)} & = & \hat{\phi}\, .
\end{eqnarray}
We get
\begin{equation}
S=\frac{1}{2}\int d^{5}x e^{-2\hat{\phi}} \sqrt{-\hat{g}} [-\hat{R}+
4 (\partial\hat{\phi})^{2} -\frac{3}{4}\hat{H}^{2}]\, .
\end{equation}
As a second step we reduce from $d=5$ to $d=4$, keeping the non-diagonal
components of the metric and 2-form field. Since we are interested also in
supersymmetry, we will work with the 5-beins at this stage. The
4-dimensional indices
do not carry a hat.
We parametrize the
$5$-beins as follows
\begin{equation}
(\hat{e}_{\hat{\mu}}{}^{\hat{a}})=
\left(
\begin{array}{cc}
e_{\mu}{}^{a} &  A_{\mu} \\
0           &    1      \\
\end{array}
\right)
\, ,
\hspace{1cm}
(\hat{e}_{\hat{a}}{}^{\hat{\mu}})=
\left(
\begin{array}{cc}
e_{a}{}^{\mu} & -A_{a} \\
0           & 1   \\
\end{array}
\right)\, ,
\label{eq:basis}
\end{equation}
where  $A_{a}=e_{a}{}^{\mu}A_{\mu}$.
With this parametrization, the 5-dimensional fields decompose as follows
\begin{eqnarray}
\hat{g}_{44} & = & \hat{\eta}_{44}=-1 \, ,
\nonumber \\
\hat{g}_{4\mu} & = & - A_{\mu}\, ,
\nonumber \\
\hat{g}_{\mu\nu} & = & g_{\mu\nu}-A_{\mu} A_{\nu}\, ,
\nonumber \\
\hat{B}_{4\mu} & = & B_{\mu}\, ,
\nonumber \\
\hat{B}_{\mu\nu} & = & B_{\mu\nu}+A_{[\mu}B_{\nu]}\, ,
\nonumber \\
\hat{\phi} & = & \phi \ ,
\end{eqnarray}
where $\{ g_{\mu\nu},B_{\mu\nu},\phi,A_{\mu},B_{\mu} \}$ are the
$4$-dimensional fields.

The $4$-dimensional action for the $4$-dimensional fields becomes.
\begin{eqnarray}
S & = &
\frac{1}{2}\int d^{4}x e^{-2\phi}\sqrt{-g} [-R
+4(\partial\phi)^{2} -\frac{3}{4}H^{2}
\nonumber \\
&  &  +\frac{1}{4} F^{2}(A)
+\frac{1}{4} F^{2}(B)]\, ,
\end{eqnarray}
where
\begin{eqnarray}
F_{\mu\nu}(A) & = & 2\partial_{[\mu}A_{\nu]}\, ,
\nonumber \\
F_{\mu\nu}(B) & = & 2\partial_{[\mu}B_{\nu]}\, ,
\nonumber \\
H_{\mu\nu\rho} & = & \partial_{[\mu}B_{\nu\rho]}+
\frac{1}{2}\{A_{[\mu}F_{\nu\rho]}(B)+B_{[\mu}F_{\nu\rho]}(A)\}\, .
\end{eqnarray}
Now,  we study the dimensional reduction of gravitino. We are specifically
interested in  the supersymmetry transformation rule
of gravitino in $d=4$ supergravity without matter. This leads to
identification of
 the
matter vector fields $D_{\mu}$ and the supergravity vector fields
$V_{\mu}$.
\begin{eqnarray}
D_{\mu} & = & \frac{1}{2}(A_{\mu}-B_{\mu})\, ,
\nonumber \\
V_{\mu} & = & \frac{1}{2}(A_{\mu}+B_{\mu})\, ,
\label{vec} \end{eqnarray}
respectively.
 Now we want to truncate the theory
keeping only the supergravity vector field $V_{\mu}$. We have then
\begin{equation}
V_{\mu}=A_{\mu}=B_{\mu}\, ,
\hspace{1cm}
D_{\mu}=0\, .
\label{eq:truncation}
\end{equation}
The truncated action is\footnote{This action, which came from  the
10-dimensional
theory is slightly different from the
corresponding 4-dimensional action in our previous papers, e.g. in \cite{US},
due to
the difference in notation. The detailed explanation of this difference will be
given in \cite{BKO2}.}
\begin{equation}
S=\frac{1}{2}\int d^{4}x
e^{-2\phi}\sqrt{-g}[-R+4(\partial\phi)^{2}-\frac{3}{4}H^{2}
+\frac{1}{2}F^{2}(V)]\, ,
\label{eq:action4trunc}
\end{equation}
where
\begin{eqnarray}
F_{\mu\nu}(V) & = & 2\partial_{[\mu}V_{\nu]}\, ,
\nonumber \\
H_{\mu\nu\rho} & = & \partial_{[\mu}B_{\nu\rho]}+
V_{[\mu}F_{\nu\rho]}(V)\, .
\end{eqnarray}

The embedding of the $4$-dimensional fields in this action in $d=10$ is
the following:
\begin{eqnarray}
g^{(10)}_{\mu\nu} & = & g_{\mu\nu}-V_{\mu}V_{\nu}\, ,
\nonumber \\
g^{(10)}_{4\nu} & = & - V_{\nu}\, ,
\nonumber \\
g^{(10)}_{44} & = & -1\, ,
\nonumber \\
g^{(10)}_{IJ} & = & \eta_{IJ}=-\delta_{IJ}\, ,
\nonumber \\
B^{(10)}_{\mu\nu} & = & B_{\mu\nu}\, ,
\nonumber \\
B^{(10)}_{4\nu} & = & V_{\nu}\, ,
\nonumber \\
\phi^{(10)} & = & \phi\, .
\label{eq:uplift}
\end{eqnarray}
This formulae can be used to uplift  any $U(1)$ $4$-dimensional field
configurations, including dilaton and axion,  to a $10$-dimensional field
configurations in a way consistent with supersymmetry.

The
conclusion of this supersymmetric dimensional reduction is the following.

i) The dilaton of the supersymmetric 4-dimensional extreme black holes is
identified as a fundamental
dilaton of string theory (and not one of the modulus fields).

ii)  Dimensional
reduction of $d=10$ supergravity to $d=4$
gives $N=4$ supergravity without 6 matter multiplets under condition that
$g_{4 \mu  }= - B_{4 \mu }$.  Therefore the
vector field of the 4-dimensional configuration is actually a non-diagonal
component of the metric in the extra
dimension as well as the 2-form field.
This works in our case since we have according to (\ref{10bh})
\begin{equation}
g_{4 t }^{(10)} = - B_{ 4 t }^{(10)} = -V_t = e^{2\phi}\ .
\end{equation}

4. We will use the formulae from the section above to uplift the
dilaton black hole with one vector field.
The
electrically charged extreme 4d black hole is given by \cite{G} \footnote{There
is a
 difference of a $1/\sqrt{2}$ factor in the vector field with
respect to the one given in \cite{US}. }.
\begin{eqnarray}
ds^{2}_{str} & = & e^{4\phi}dt^{2}-d\vec{x}^{2}\, ,
\nonumber \\
V & = &  - e^{2\phi}dt\, ,
\nonumber \\
B & = & 0\, ,
\nonumber \\
e^{-2\phi} & = & 1+\frac{2M}{\rho}\, .
\end{eqnarray}
The uplifted configuration, according to eq. (\ref{eq:uplift}) is:
\begin{eqnarray}
ds^{2} & = & 2e^{2\phi}dtdx^{4}-d\vec{x}^{2}-(dx^{4})^{2}-
dx^{I}dx^{I}\, ,
\nonumber \\
B^{(10)} & \equiv & B^{(10)}_{MN}dx^{M}\ wedge dx^{N}=- 2e^{2\phi}dx^4\wedge
dt \, ,
\nonumber \\
\phi^{(10)} & = & \phi\, .
\label{upl}\end{eqnarray}

Let us choose the parameter $\mu$ in the dual partner to the wave,
given in eq.  (\ref{partner}) equal to the double mass of the black hole.
\begin{equation}
\mu = 2 M\ .
\end{equation}
This makes the uplifted black hole (\ref{upl}) identical to the dual
partner to the wave,
given in eq.  (\ref{partner}).

For better understanding of black-hole-wave relation it is useful to do the
following.
By adding and subtracting from the metric the term
$e^{4\phi} dt^2$  we can rewrite  the dual wave in $d=10$, given
in eq. (\ref{10bh}) as follows.
\begin{eqnarray}
\label{eq:bh}
ds^2 &=&  e^{4\phi} d t ^2 -
 d \vec x^2 - (dx^4  -
 e^{2\phi} dt )^2
 -  dx^Idx^I
\ ,\nonumber\\
B &=&
  - 2e^{2\phi}dx^4\wedge
dt  \ ,\\
e^{-2\phi}  &=& 1 +  { 2M \over \rho}\ .\nonumber
\end{eqnarray}
 Now it is easy to recognize in the first 2 terms in the metric the
4-dimensional
metric and in the third term the non-diagonal component of the
10-dimensional metric  which together with the non-diagonal component of
the 2-form plays the role of the vector field in the 4-dimensional geometry.

5.   The case $\xi^2 = -1$ which gives  the 4-dimensional black hole in
Minkowski space with the signature  $(1,3)$
times the
compact 6-dimensional space with the signature $(0,6)$ corresponds to a
complex
10-dimensional wave in the space
with the signature  $(1,9)$.
\begin{eqnarray}
ds^2 &=& 2d\tilde u d \tilde v -  \; { 4M \over \rho}\;d\tilde
u( d\tilde u -i d\tilde x^4) - \sum_{i=1}^{i=8}
 d\tilde x^id\tilde x^i\ ,\nonumber \\
B &=& -i \;{ 4M \over \rho}\; d \tilde u \wedge d\tilde x^4\ .
\label{wave} \end{eqnarray}
By  performing a  rotation $i\tilde x^4 = \tilde \tau$  one can get
\begin{eqnarray}
ds^2 &=& 2d\tilde u d \tilde v  -  \; { 4M \over \rho }\;d\tilde u( d\tilde u
-d\tilde \tau)+ d\tilde \tau^2  -
\sum_{i=1}^{i=7} d\tilde x^i d \tilde x^i\ ,\nonumber \\
B &=&  - \;{ 4M \over \rho }\; d\tilde u \wedge d\tilde \tau\ \ .
\label{Brink}
\end{eqnarray}
This makes the wave real but with the signature of the space $(2,8)$.

Thus we may conclude that string theory considers as dual partners the
extreme 4d electrically
charged dilaton black hole  embedded into 10-dimensional geometry,
as given in eq. (\ref{eq:bh}) or (\ref{upl}),
and Brinkmann-type
10-dimensional  wave (\ref{wave}), (\ref{Brink}).

If we choose $\xi^2 = 1$ case we get the stringy equivalence between
Brinkmann-type 10-dimensional wave
\begin{eqnarray}
ds^2 &=& 2d\tilde ud \tilde v -  \; { 4M \over \rho}\;d\tilde
u( d\tilde u - d\tilde x^4) - \sum_{i=1}^{i=8}
 d\tilde x^id\tilde x^i\ ,\nonumber \\
B &=& -\;{ 4M \over \rho}\; d \tilde u \wedge d\tilde x^4\ ,
\label{SSW2} \end{eqnarray}
and lifted Euclidean 4-dimensional electrically charged dilaton
black hole with the signature
$(0,4)$ and
the 6-dimensional space has the signature  $(1,5)$,
\begin{eqnarray}
\label{eq:euclbh}
ds^2 &=& -e^{4\phi} d t ^2 -
d \vec x^2 + (dx^4 +
 e^{2\phi} dt)^2
- dx^Idx^I
\ ,\nonumber\\
B &=&
 - 2 e^{2\phi}dx^4 \wedge dt   \ ,\nonumber\\
e^{-2\phi}  &=& 1 +  { 2M \over \rho}\ .
\end{eqnarray}
With such choice of the signature the gravitational wave does not
have imaginary components. However,
the  fact that the metric as well as the 2-form field of the
gravitational wave in $d=10$ have an imaginary component  to be
dual to the lifted black hole in Minkowski space is strange. Note that this
is necessary only
if one insists that the $d=10$ space as well as the $d=4$ space are both
Minkowski spaces. One can avoid imaginary components by allowing
the changes in the signature of the space-time when performing duality
and dimensional reduction as explained above. Still this remains a puzzle.

 Could we actually consider the dual relation between waves and
lifted black holes as something more than pure  algebraic curiosity?
 We believe that the answer to this question is ``yes''.  The dual relation
displayed above was established at the zero slope limit of the effective
action of the superstring theory. The issue of $\alpha' $-corrections in string
theory has
been studied extensively for the waves \cite{BKO}, \cite {BEK}. The pp-waves
have the best known
properties of absence of such quantum corrections \cite {Gu1}. The SSW  are
known to
have special property of the absence of $\alpha' $-corrections under the
condition that the non-abelian Yang-Mills fields  is added to the configuration
which at the zero slope limit $\alpha' =0$ consists only of the metric and
2-form \cite{BKO}.
It was explained in \cite{BEK} that the importance of
sigma-model duality  between supersymmetric configurations is in the
fact that the structure of
$\alpha'$ corrections is under control for the
dual solution if it was under control for the original solution.
In this way we have found that the nice properties of the pp-waves\cite{Gu1}
are
carried over to the fundamental string solutions. The present investigation
shows that
the electrically charged extreme black hole embedded into 10-dimensional
geometry may require to be supplemented by some non-abelian Yang-Mills
configuration  to avoid the possible $\alpha' $-corrections. In this respect
we would like to stress that the study of the properties of quantum corrections
established via duality may become a powerful mechanism of the
 investigation of  quantum theory despite the strange imaginary factors
in the waves, which are dual partners of the uplifted black holes.

At the very minimal level one can consider the method developed above,
which consists of stringy duality combined with Kaluza-Klein dimensional
reduction, as the
solution generating method. This method has the advantage of generating new
supersymmetric solutions from the original ones. If  we did not know that
 extreme
4-dimensional black holes are supersymmetric, we would discover this
via the supersymmetric properties of 10-dimensional gravitational waves. We
hope
to
derive more general 4-dimensional supersymmetric solutions starting from
our 10-dimensional supersymmetric waves and to explore generic relation
between supersymmetry and duality \cite{BKO2}.

We are grateful to G. W. Gibbons, G. T. Horowitz,  J. H. Schwarz and A.
Tseytlin for
 the most fruitful discussions.
We are extremely grateful to the
 organizers of the programme ``Geometry and
Gravity"
at the Newton Institute, Cambridge, which has allowed us to come and work
together. One of us (T.O.) would like to thank Groningen
University for the hospitality.
The work of E.B. and  R.K. was partially supported by a NATO
Collaborative Research Grant. The work of E.B. has been made
possible by a
fellowship of the Royal Netherlands Academy of Arts and Sciences
(KNAW).
The work of R. K. was  supported by
NSF
grant PHY-8612280 and the work of T. O. was supported by  European
Communities Human Capital and Mobility programme grant.

\end{document}